\documentclass[10pt,conference]{IEEEtran}
\IEEEoverridecommandlockouts
\usepackage{algorithmic,algorithm}
\usepackage[left=0.75in,right=0.75in,top= 0.85in,bottom=1.3in]{geometry}
\usepackage{amsmath}
\usepackage{textcomp}
\usepackage{amssymb,amsmath}
\usepackage{graphicx}
\usepackage{epsfig}
\usepackage{grffile}
\usepackage{balance}
\usepackage{balance,color}
\usepackage{algorithmic,algorithm}
\providecommand{\keywords}[1]{{\textit{Index Terms}}}

\def\BibTeX{{\rm B\kern-.05em{\sc i\kern-.025em b}\kern-.08em
    T\kern-.1667em\lower.7ex\hbox{E}\kern-.125emX}}
		
\begin{document}

\title{ Compress or Interfere? }
\author{ Alaa Awad Abdellatif$^{*\dag}$, Lutfi Samara$^{+}$, Amr Mohamed$^{*}$, Abdulla Al-Ali$^{*}$, Aiman Erbad$^{*}$,\\ and  Mohsen Guizani$^{*}$  \\
$^*$Department of Computer Science and Engineering, Qatar University, Qatar \\
$^+$Department of Electrical Engineering, Qatar University, Qatar \\
$^\dag$Department of Electronics and Telecommunications, Politecnico di Torino, Italy  \\
E-mail: alaa.abdellatif@polito.it, \{samara, amrm, abdulla.alali, aerbad, mguizani\}@qu.edu.qa  \\ 
\thanks{978-1-7281-2294-6/19/\$31.00 ©2019 IEEE}
}
\maketitle


\begin{abstract}
Rapid evolution of wireless medical devices and network technologies has fostered the growth of remote monitoring systems. Such new technologies enable monitoring patients' medical records anytime and anywhere without limiting patients' activities. However, critical challenges have emerged with remote monitoring systems due to the enormous amount of generated data that need to be efficiently processed and wirelessly transmitted to the service providers in time. Thus, in this paper, we leverage full-duplex capabilities for fast transmission, while tackling the trade-off between Quality of Service (QoS) requirements and consequent self-interference (SI) for efficient remote monitoring healthcare systems. 
The proposed framework jointly considers the residual SI resulting from simultaneous transmission and reception along with the compressibility feature of medical data in order to optimize the data transmission over wireless channels, while maintaining the application's QoS constraint.  
 Our simulation results demonstrate the efficiency of the proposed solution in terms of minimizing the transmission power, residual self-interference, and encoding distortion.    
\end{abstract}

\section{Introduction\label{sec:Introduction}}
 
 The prompt growth in wireless body sensor network, Internet of Things (IoT), and wireless network technologies has accelerated implementing efficient remote healthcare services \cite{HERMIT}, \cite{Survey_IMD}. Leveraging such emerging technologies enables the healthcare service providers to continuously monitor large number of patients without visiting the hospitals. This evaluation significantly helps in a variety of pre-hospital emergency situations \cite{SmartHealth19}.   
However, transferring the massive amount of data generated from such systems is challenging. 
For example, in Intensive Care Unit of Electroencephalography (EEG) monitoring system, samples of EEG signal along with video recording should be stored and accessed remotely for correlating clinical activity with EEG pattern, which leads to generating 8-10 GB per patient every day \cite{ContinuousMonitoring}. This large amount of generated data obviously sets a significant load on the system performance and scalability in terms of processing capabilities, storage, transmission power, and wireless resources \cite{AlaaTrns2018}.    

Thus, we envision that Full-duplex (FD) communications can play a significant role to meet this demand for high data rates. FD communications has enabled a wireless transceiver to simultaneously transmit and receive wireless information on the same frequency band, hence potentially doubling the system sum-rate compared with wireless Half-Duplex (HD)  transmission \cite{syrjala2014analysis}, \cite{samara2017residual}, \cite{doublerate}. 
However, FD communication cannot be efficiently exploited without dealing with the self-interference (SI) coming from the simultaneous transmission and reception at the same band and time. Recently, many related works have investigated the possibility of leveraging FD transmission in wireless communications by reducing the SI \cite{FullHalf2018}, \cite{FullHalf2011}. 
 FD bidirectional communication was demonstrated to be possible if the wireless transceiver possesses SI cancellation capabilities \cite{korpi2017compact}, \cite{choi2010achieving}. In FD transceivers, the SI should be suppressed as much as possible to reach the noise level in order to achieve a reliable communication. However, the residual SI is the performance bottleneck that hinders the FD transceiver from realizing its potential capabilities, hence it must be closely investigated. Several articles have considered the effect of the residual SI on the performance of various communication systems. For instance, buffer-aided FD relaying networks are studied under the presence of residual SI, where the residual SI was modeled to be proportional to the relay’s transmission power \cite{phan2015buffer}. In \cite{cheng2013optimal}, the residual SI was again considered in developing optimal dynamic power control algorithms in a FD bidirectional communication link. These studies along many others highlight the importance of the residual SI in analyzing the performance of a FD communication system. However, none of the aforementioned works have considered the effect of optimizing both the generated data rate and transmission power consumption in order to minimize the residual SI without violating application's QoS requirements.    

In this paper, we argue that leveraging adaptive compression techniques along with FD communications can significantly improve system performance and emergency response time for remote health monitoring applications.   
Thus, the main contributions of this paper can be highlighted as follows.
\begin{enumerate}
	\item  Formulate a multi-objective optimization problem that enables the Patient/personal Data Aggregator (PDA) to optimally adjust its transmission rate, such that it guarantees a minimum level of residual SI, while ensuring an acceptable distortion level.
	\item  Derive analytically a closed-form solution for the formulated optimization problem. We remark that the proposed closed-form solution allows for a quick and simple calculation for the optimal transmission rate and compression ratio, thus it is amenable for implementation at the PDA (i.e., mobile edge).
	\item The proposed solution is evaluated through simulations discussing the tradeoff between residual SI and encoding distortion. Our results demonstrate the effectiveness of our solution and its ability to outperform the conventional wireless half-duplex transmission	schemes. 
\end{enumerate}

The rest of the paper is organized as follows. Section II describes the system model. Section III presents the adopted performance metrics and the proposed optimization problem formulation. Section IV introduces the analytical solution of the proposed problem. Section V presents the stimulation results, while Section VI concluding the paper. 
\section{System Model \label{sec:system}}

\begin{figure}[t!]
	\centering
		\scalebox{1.8}{\includegraphics[width=0.27 \textwidth]{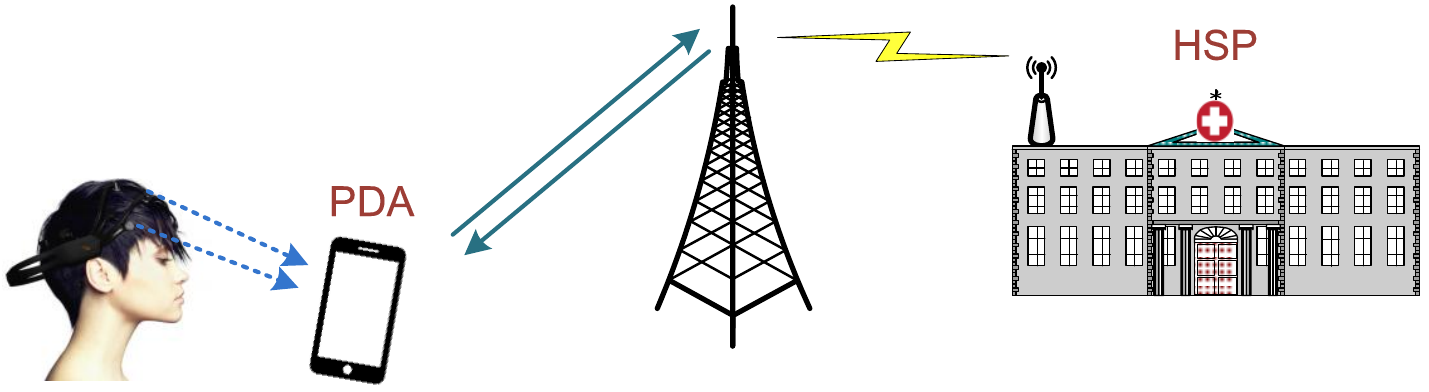}}
	\caption{System model under study. }
	\label{fig:System}
\end{figure}

In this paper, the remote monitoring system shown in Figure \ref{fig:System} is investigated, where we consider the full-duplex transmission over bidirectional wireless channels with imperfect self-interference cancellation. In particular, the proposed  framework exploits the electroencephalogram (EEG) dataset in \cite{EEGdataset}, which has been widely used to study the electrical activities of the brain. Intensive brain monitoring has a significant role in treating many types of brain disorders, such as Parkinson's disease and epilepsy \cite{brain_stimulation_epilepsy}. However, without loss of generality, the proposed framework can be easily extended to consider different biosignals or video streaming applications.

 The proposed network topology consists of a PDA that collects the EEG data from a patient. Then, it compresses and forwards the compressed data to the Healthcare Service Provider (HSP) using FD communication. HSP is responsible for reconstructing and analyzing the received data to obtain the patient's state, then sending the required feedback/services to the patient.        
Thus, for implementing a swift and efficient bidirectional remote monitoring system, we propose the following tasks at the PDA:
\begin{enumerate}
	\item Optimizing both the encoding distortion and transmission power, while considering the channel characteristics and application's requirements in order to obtain the optimal compression ratio and transmission rate. We remark here that the residual SI is typically proportional to the transmission power \cite{cheng2013optimal}. Also, increasing the transmission power indefinitely will not necessary result in a maximized data rate due to the effect of residual SI.  Hence, we can optimize the residual SI by minimizing the transmission power, as will be shown later.  
	\item Compressing acquired medical data using the obtained compression ratio.
	\item Transmitting compressed data using FD communication.
\end{enumerate}
 Accordingly, the proposed optimization model enables the PDA to adaptively reconfigure the transmitted data rate and transmission power, hence minimizing the residual SI, while retrieving the original data at the HSP. Thus, at the HSP, data reconstruction, storage, and further sophisticated analysis can be executed to evaluate the patient's state. 
\section{Performance Metrics and Problem Formulation \label{sec:Metrics}}
In what follows, we obtain the QoS requirement of our healthcare application, which includes the encoding distortion, and the required transmission power to forward the generated data. Then, we propose our Power-Distortion optimization model.  

\subsection{Performance Metrics \label{sec:Metrics}}

The required transmission rate of a FD transceiver over a wireless link with bandwidth $B$ is defined as a function of the signal-to-interference-plus-noise ratio (SINR) as follows:
\begin{equation}
	R = B\log_2(1+SINR) = B\log_2\left(1+\frac{h \cdot P}{N_0 + \mu \cdot P}\right),  
 \label{eq:Rate}
\end{equation}
where $h$ is the channel power gain that is modeled as a random variable following Rayleigh distribution, $N_0$ is the noise spectral density, and $\mu$ is a constant that reflects the quality of the employed SI cancellation scheme\footnote{Smaller values of $\mu$ represents a better SI cancellation quality and vice versa.} \cite{cheng2013optimal, wong_schober_ng_wang_2017}. In (\ref{eq:Rate}), a bidirectional communication model is considered with a fixed transmission power $P$. Thus, with straightforward manipulations,  the transmission power $P$ can be defined as
 \begin{equation}
	P = \frac{N_0\left(2^{\frac{R}{B}} -1 \right)}{h - \mu\left (2^{\frac{R}{B}} -1 \right )}.   
 \label{eq:power}
\end{equation} 

\begin{figure}[t!]
	\centering
		\scalebox{1.7}{\includegraphics[width=0.27 \textwidth]{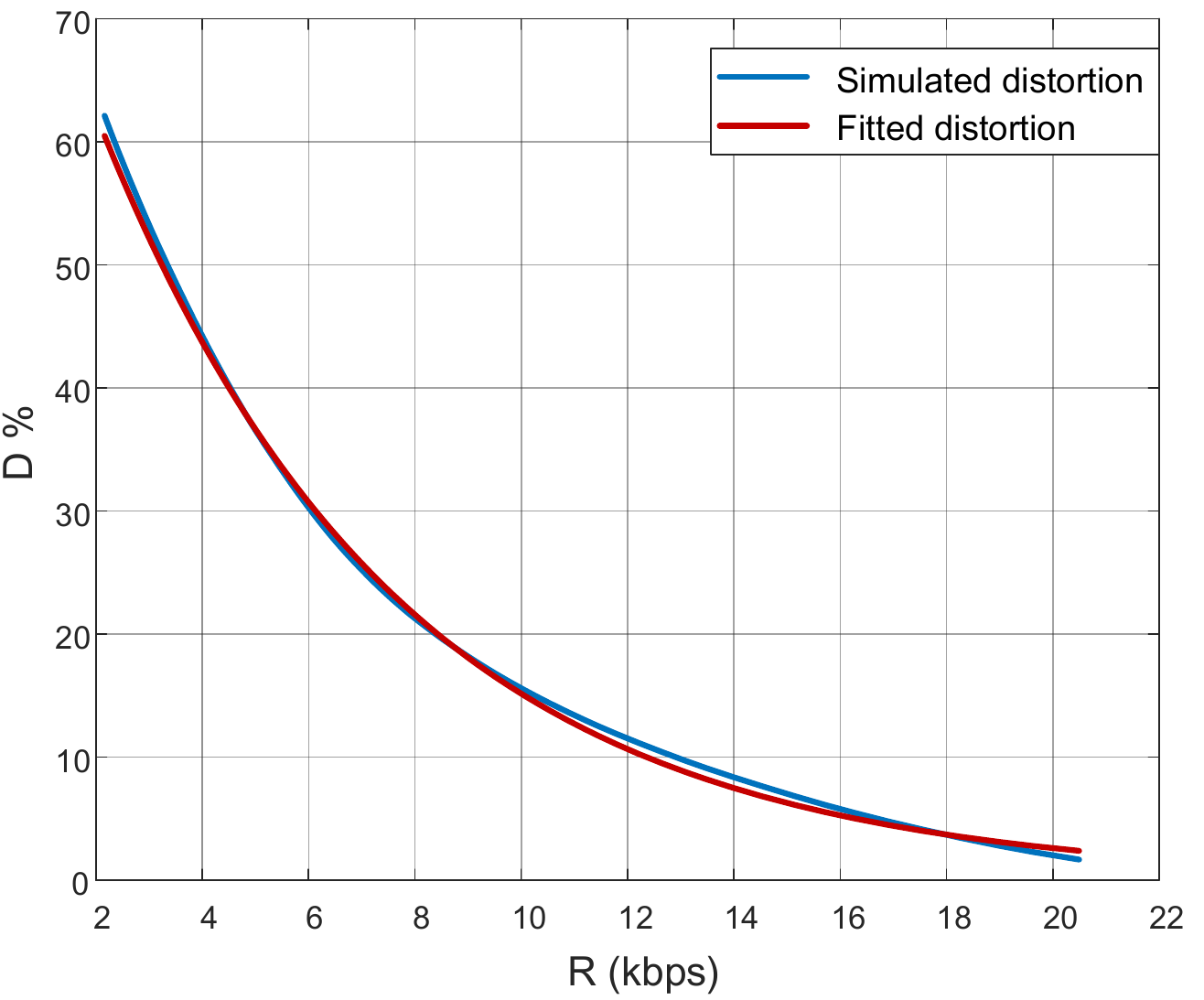}}
	\caption{Comparison between the simulated distortion and the analytical expression in (\ref{eq:des}). }
	\label{fig:Dis_rate_fit}
\end{figure}

As mentioned earlier, the residual SI of a FD transceiver is proportional to the transmission power \cite{cheng2013optimal}, hence, to minimize the residual SI, we have to optimize the transmission power.     
Thus, we opt to leverage an adaptive compression scheme in order to adjust the generated data rate based on the transmission power and characteristics of the wireless channels. Hence, our scheme proposes an extra degree of freedom for the PDA to transfer the acquired medical data without violating the application's QoS and residual SI constraints.   
To achieve that,  the PDA compresses the EEG signals using threshold-based Discrete Wavelet Transform (DWT) \cite{implementationpaper}. In DWT, the EEG signal $x$ is written as:
$	x = \Psi \alpha_w$,    
 where $\Psi$ is the Daubechies wavelet family basis, and $\alpha_w$ is the vector of wavelet domain coefficients. 
Hence, DWT coefficients that are below a predefined threshold $\delta$ can be zeroed and discarded from transmission without much encoding distortion. Consequently, by properly selecting the threshold $\delta$, we can adjust the length of the transmitted data, hence the compression ratio $\kappa$ of our scheme.   
At the receiver side, the reconstruction and data recovery are performed using inverse DWT to retrieve the original data. However, the reduction in the transmitted data size comes with a distortion overhead caused by the lossy compression. This distortion is measured  by  the percentage  root-mean-square  difference  between the recovered EEG data and the original one, as 
\begin{equation}
	  D = \frac {\| {\textbf{x}} - \hat{{\textbf{x}}} \|}{\| {\textbf{x}} \|} \cdot 100 , 
	\end{equation}
where $\textbf{x}$ and $\hat{\textbf{x}}$ are the vectors of the original and the reconstructed signals, respectively, and $\| .\|$ represents the norm operator. To optimize this distortion in our framework, an analytical model is derived to quantify the distortion using data regression techniques. The obtained analytical model of the distortion is written as a function of the generated data rate $R_s$ as follows 
\begin{equation} 
	  D = a\exp(b \cdot R_s), 
 	\label{eq:des}
\end{equation}
where the function $\exp()$ represents the exponential function operator, and $a$ and $b$ represent the model parameters, which are estimated based on the statistics of the used EEG data and DWT encoder. It is worth mentioning here that the above regression model results in a coefficient of determination of around 0.99.  Figure \ref{fig:Dis_rate_fit} depicts the simulated and fitted distortions as modeled in (\ref{eq:des}). 
Thus, the generated data rate $R_s$ turns to be a function of the compression ratio $\kappa$, the sampling frequency of the acquired data $f_s$, and the number of bits per sample $n$ as follows 
\begin{align}
R_s = n\cdot f_s \cdot (1-\kappa).
\end{align}
We remark here that different data types with different lossy compression techniques can be considered within the proposed framework by modifying/replacing the distortion model in (\ref{eq:des}).  

The main question now is: How can we obtain the optimal data rate $R$ to guarantee an efficient transmission of our medical data, while optimizing the trade-off between the consequent distortion and residual SI? We answer this question in the following section.        
 	
\subsection{Problem Formulation  \label{sec:Formulation}}

The proposed optimization problem considers two conflicting objectives: residual SI minimization and distortion minimization. It is clear from (\ref{eq:power}) that increasing the transmission rate $R$ leads to increasing the transmission power, hence increasing the residual SI. Also, to decrease the transmission rate using lossy compression techniques (by decreasing the generated data rate $R_s$) the compression ratio increases, which leads to increasing the distortion. Hence, it is a crucial to obtain the optimal trade-off between the distortion and transmission power by precisely defining the optimal compression ratio and transmission rate. To consummate this, we define a single aggregate objective function, which turns the above conflicting objectives into a single objective function.   
However, each objective presents different ranges and units of measurement, hence they need first to be normalized. The goal of the normalization process is to map these objective onto non-dimensional values within the [0,1] range to make them comparable \cite{ALaa2017Elsevier}. Therefore, the optimization problem is formulated as follows: 
\begin{eqnarray}
\mbox{\bf P:} &	&   \min_{R, \kappa} \   \lambda \cdot \tilde{P} +(1-\lambda) \cdot \tilde{D}  
	 \label{eq:optimize_prob} \\
\mbox{s.t.} &&	R_s \le R < B\log_2(1+{h}/{\mu})   \label{eq:C1} \\
	&& 0 \le \kappa \le 1  \label{eq:C2} \\
	&& 0 \le \lambda \le 1. \label{eq:C3}
\end{eqnarray}
In this optimization problem: 
\begin{itemize}
	\item $\tilde{P} = \frac{P}{P_{max}}$ is the normalized transmission power, where $P_{max}$ is the maximum transmitted power.
	\item $\tilde{D}=\frac{\log(D)}{\beta}$ is the normalized logarithmic function of the distortion, where $\beta$ is a normalization factor. Herein, we consider $\tilde{D}$ because it is turned to be a linear function in the data rate $R_s$, which enables us to obtain a closed-form solution for the proposed optimization.  
	\item $\lambda$ is the weighting coefficient which represents the relative importance of the two objective functions in our problem.  
	\item The flow rate constraint is represented by (\ref{eq:C1}) to guarantee that the whole generated data is transmitted, where $R_s$ is the generated data rate. It is worth pointing here that, for the power feasibility,
we must have $R_s < B\log_2(1+{h}/{\mu})$.       
\end{itemize}
The formulated objective function in (\ref{eq:optimize_prob}) is not a convex function with respect to $R$, since $\frac{\partial{}^2}{\partial{R}^2}(\tilde{P})<0$  at $R>>B$. Hence, it cannot be solved using well-known convex optimization tools \cite{a15}. 

\section{Power-Distortion Optimization \label{sec:Optimization}}
In this section, we envision a methodology to simplify the proposed optimization problem in (\ref{eq:optimize_prob}) such that an optimal, analytical solution is obtained.

\textbf{Theorem:} \textit{In the proposed optimization problem in (\ref{eq:optimize_prob}), the optimal transmission rate and compression ratio of a PDA with data rate $R_s$ is given as}
\begin{equation}
R^* = B\log_2(\chi), \  \  \kappa^* = \frac{1-B\log_2(\chi)}{f_s \cdot n}.  \nonumber\\
\end{equation} 
where
{\small  
\begin{equation}
\chi = \frac{-(2\tilde{b} B \mu c_2 + c_1) + \sqrt{(2\tilde{b} B \mu c_2 + c_1)^2 - 4\tilde{b}^2 B^2 \mu^2 {c_2}^2}}{2\tilde{b} B \mu^2}.
\label{eq:chi}
\end{equation} 
\normalsize}

\begin{IEEEproof} 
{
Looking at problem formulation in (\ref{eq:optimize_prob}), one can see that optimal transmission rate $R$ can be only achieved at $R = R_s = n\cdot f_s \cdot (1-\kappa)$, since increasing $R$ leads always to a significant increase in $\tilde{P}$. Also, $R$ cannot decrease below $R_s$ in order to guarantee the stability of the system. Accordingly, our objective function (i.e., $U = \lambda \cdot \tilde{P} +(1-\lambda) \cdot \tilde{D} $) turns out to only be a function of $R$, which depends on the adopted compression ratio $\kappa$. Thus, the unknown in our problem now is $R$, i.e., each PDA needs to determine its transmission rate based on the optimal $\kappa$, in order to maintain the optimal trade-off between its transmission power and distortion.  
Accordingly, a closed-form expression for the optimal $R$ and $\kappa$ can be obtained by imposing that the derivative with respect to $R$ of the objective function $U= \lambda \cdot \tilde{P} +(1-\lambda) \cdot \tilde{D} $ is equal to $0$. Thus,

{\small 
\begin{eqnarray} 
&\frac{\partial{U}}{\partial{R}} & = \frac{\partial{}}{\partial{R}} \left[ \lambda \cdot \tilde{P} + (1-\lambda)\cdot \tilde{D} \right ] \nonumber\\
& = &\frac{\partial{}}{\partial{R}}\left [ \frac{\frac{\lambda \cdot N_0}{P_{max}}\left(2^{R/B}-1\right)}{h-\mu \left(2^{R/B}-1\right)}  + \frac{(1-\lambda)\log(a\exp(bR))}{\beta} \right ] \nonumber\\
& = &\frac{\partial{}}{\partial{R}}\left [ \frac{\tilde{N}\left(2^{R/B}-1\right)}{h-\mu \left(2^{R/B}-1\right)}  + \frac{(1-\lambda)(\log(a)+bR)}{\beta} \right ] \nonumber\\
& = & \frac{\log(2)\cdot h \cdot \tilde{N} \cdot 2^{R/B}}{B(\mu \cdot 2^{R/B}-h-\mu)^2}  + \tilde{b} =0,   \nonumber\\
\label{derivative}
\end{eqnarray} 
\normalsize}
where $\tilde{N}=\frac{N_0 \cdot \lambda}{P_{max}}$, and $\tilde{b}=\frac{b(1-\lambda)}{\beta}$.  
For the sake of clarity, let us define the following,  
\begin{align} \label{eq:defins}
\log(2)\cdot h \cdot \tilde{N} = c_1, \  \   -h-\mu = c_2, \  \   2^{R/B} = \chi.
\end{align}
Hence, we will have
\begin{eqnarray} 
0 &=&  \frac{c_1 \cdot \chi}{B(\mu \cdot \chi - c_2)^2}  + \tilde{b}  \nonumber\\
0 &=&  (\tilde{b} B \mu^2)\chi^2 + (2\tilde{b} B \mu c_2 + c_1)\chi + \tilde{b} B {c_2}^2. \nonumber\\
\end{eqnarray} 
Hence, 
{\small 
\begin{equation} \label{eq:eqresult}
\chi=\frac{-(2\tilde{b} B \mu c_2 + c_1) + \sqrt{(2\tilde{b} B \mu c_2 + c_1)^2 - 4\tilde{b}^2 B^2 \mu^2 {c_2}^2}}{2\tilde{b} B \mu^2}.
\end{equation}
\normalsize}
By substituting from (\ref{eq:defins}) in (\ref{eq:eqresult}), we get the optimal transmission rate and compression ratio as follows
\begin{equation} \label{eq:Ropt}
\   R^* = B\log_2(\chi).
\end{equation}
\begin{equation} \label{eq:copt}
\kappa^* = \frac{1-R^*}{f_s \cdot n} = \frac{1-B\log_2(\chi)}{f_s \cdot n}. \  \  \  \  \  \ 
\end{equation}
}
\end{IEEEproof}

We remark here that leveraging the proposed scheme with adaptive compression allows for the PDA to reconfigure its compression ratio based on the characteristics of the wireless channel and application's requirements. Hence, the PDA can obtain the optimal transmission rate that minimizes both the residual SI and distortion.

\section{Simulation Results \label{sec:simulation}}

The simulation results were generated using the EEG dataset in \cite{a15}. To model small scale channel variations, flat Rayleigh fading is used with Doppler frequency 0.1 Hz and sampling time 0.1 sec. The detailed simulation parameters are reported in Table \ref{tab:SysParameters}.  
\begin{table}[htp]
	\centering
		\caption{Simulation Parameters}
	  \label{tab:SysParameters}
		\begin{tabular}{|c|c||c|c|} 
			\hline 
		\textbf{Parameter} & \textbf{Value} & \textbf{Parameter} & \textbf{Value}\\
		\hline
		$f_s$ &	\phantom {$c^{1^v}$} 2 kHz & $n$ &	\phantom {$c^{1^v}$} 12 bps \\
		\hline
	 	 $B$ & \phantom {$c^{1^v}$} 30 kHz & $N_0$ & \phantom {$c^{1^v}$} -174 dBm \\
		\hline
		$\mu$ & \phantom {$c^{1^v}$} 0.001 & $\lambda$ &	\phantom {$c^{1^v}$} 0.5   \\	
		\hline
		$a$ &	\phantom {$c^{1^v}$} 88.63	& 	$b$ &	\phantom {$c^{1^v}$} -0.0001767	 \\
	\hline
		\end{tabular}
\end{table}     

\begin{figure}[htp]
	\centering
		\scalebox{1.34}{\includegraphics[width=0.35 \textwidth]{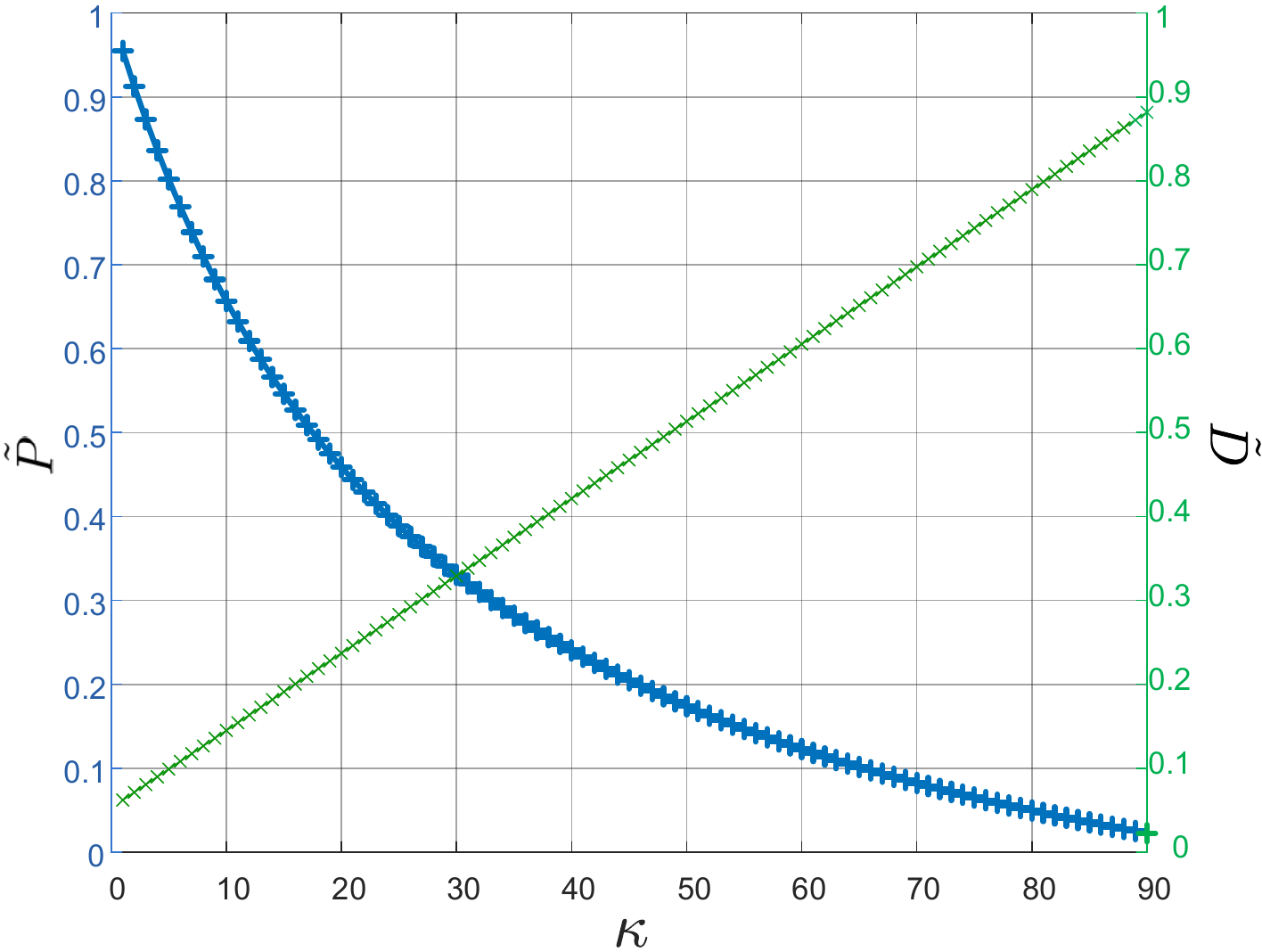}}
	\centering
	\caption{The trade-off between normalized transmission power and $\tilde{D}$ with varying compression ratio.}
	\label{fig:tradeoff}
\end{figure}

First, we display in Figure \ref{fig:tradeoff} the trade-off of between the transmission power and distortion using adaptive lossy compression technique. It is clear that by increasing the compression ratio $\kappa$, the transmission rate $R$ decreases, which result in decreasing the transmission power. However, this trend comes at the expenses of increasing the distortion. Thus, it is important to obtain the optimal value of $\kappa$ that maintains the optimal trade-off between the transmission power and distortion.

\begin{figure}[htp]
	\centering
		\scalebox{1.3}{\includegraphics[width=0.35 \textwidth]{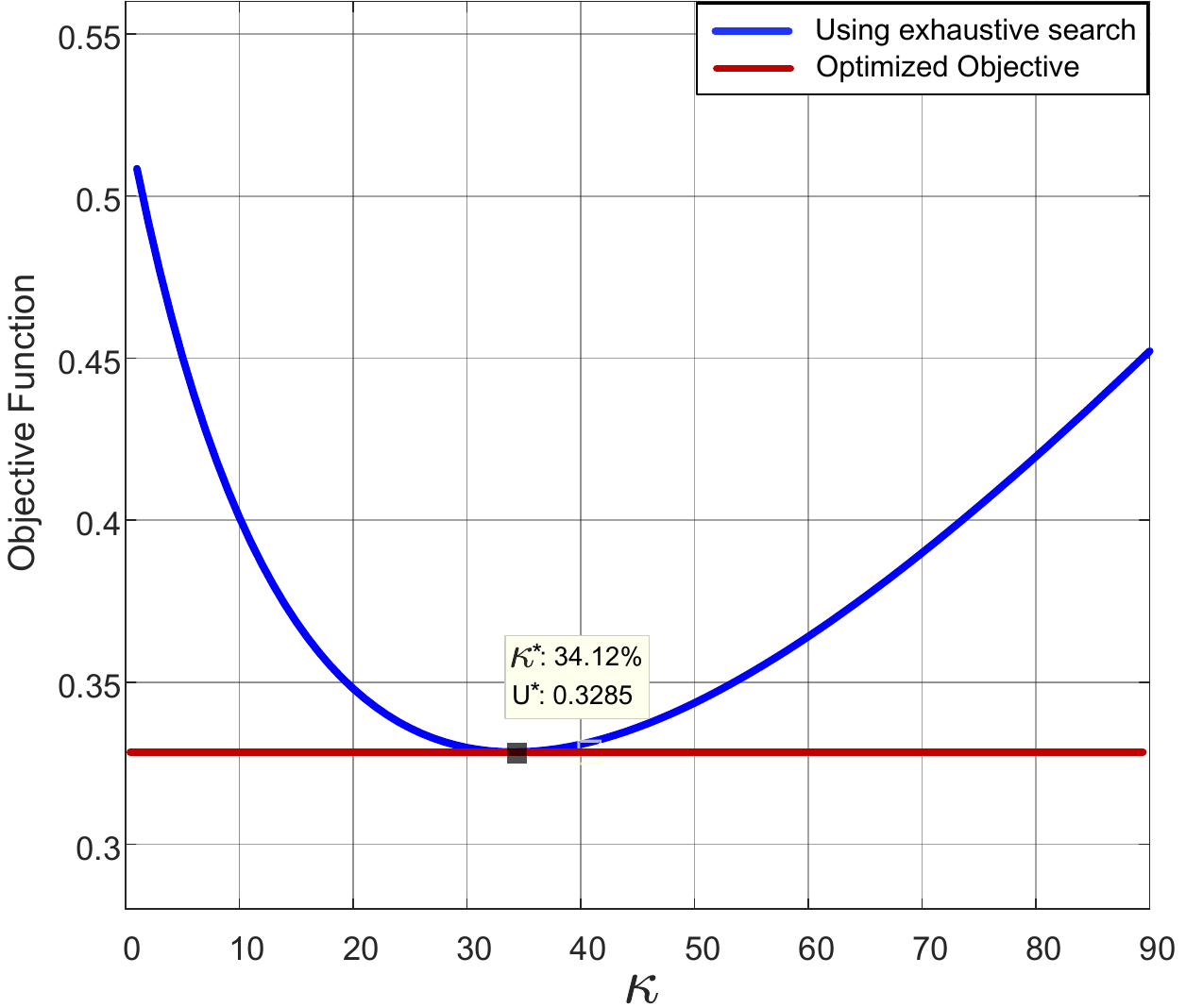}}
	\centering
	\caption{Comparison between the simulated objective function using exhaustive search and the optimized objective $U^*$.}
	\label{fig:Utility}
\end{figure}

 Figure~\ref{fig:Utility} depicts the obtained optimal compression ratio $\kappa^*$ and the objective function $U^*$ compared to the obtained values using exhaustive search. As shown, the obtained optimal $\kappa^*$ and $U^*$ using derived closed-form expression in (\ref{eq:copt}) are extremely accurate compared to the obtained solution using exhaustive search. Thus, the PDA can accurately reconfigure its compression parameters based on the wireless channel characteristics, residual SI, and application requirements in order to minimize its objective function without adding much complexity.       

\begin{figure}[t!]
	\centering
		\scalebox{1.3}{\includegraphics[width=0.35 \textwidth]{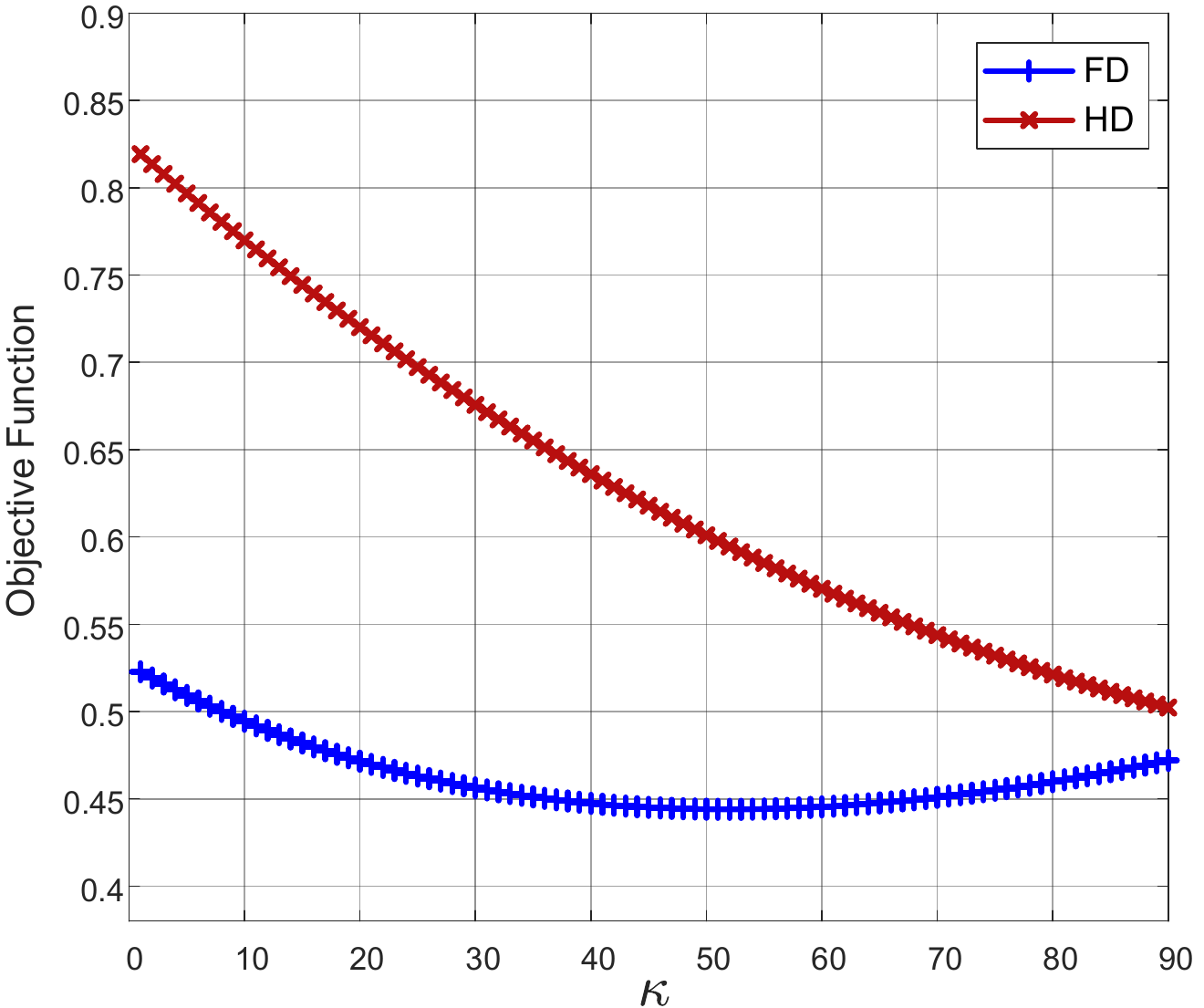}}
	\centering
	\caption{Comparison between the proposed FD solution and HD solution with varying $\kappa$.}
	\label{fig:compare}
\end{figure}

Finally, Figure~\ref{fig:compare} illustrates the main advantage of the proposed scheme compared to Half-Duplex (HD) approach. In HD, the allocated bandwidth/resources is divided between transmission and reception, hence, the allocated resources is reduced compared to FD approach. On the contrary, FD leverages all resources in both transmission and reception, at the expenses of generating residual SI.   
 Thus, adopting adaptive compression along with FD transmission, while optimizing the transmission power as a function of compression ratio could always maintain better performance compared to the HD approach (as shown in Figure~\ref{fig:compare}).  

\section{Conclusion\label{sec:conclusion}}

In this paper,  we investigate the advantages of leveraging full-duplex communications with adaptive compression techniques at the network edge for enabling efficient transmission of medical data without violating application's QoS requirements. A multi-objective optimization problem is formulated to minimize the transmission power, hence minimizing the residual self-interference, as well as the encoding distortion. Thus, our study is leveraged to identify the optimal trade-off between transmission power and distortion, based on the application's requirements. In particular, the proposed scheme is performed at the PDA level in order to optimize the transmission power as a function of the transmission rate, wireless channel characteristics, and distortion requirement. 
In this context, we could derive an analytical solution for the formulated optimization problem. Our simulation results show that the proposed optimization framework with adaptive data compression provides the optimal trade-off between the adopted conflicting objectives (i.e., transmission power and distortion), so that it outperforms the conventional half-duplex transmission techniques. 

\balance
\section*{Acknowledgment}
{
This publication was made possible by NPRP grant \# NPRP8-408-2-172 from the Qatar National Research Fund (a member of Qatar Foundation). The statements made herein are solely the responsibility of the authors. 
}
\bibliographystyle{IEEEtrannames}
\bibliography{full_duplex}   

\end{document}